# Pancreatic Cancer ROSE Image Classification Based on Multiple Instance Learning with Shuffle Instances

Tianyi Zhang[1], Youdan Feng[1], Yunlu Feng[2], Guanglei Zhang[1*]

[1] Beijing Advanced Innovation Center for Biomedical Engineering, School of Biological Science and Medical Engineering, Beihang University, Beijing, China
[2] Department of Gastroenterology, Peking Union Medical College Hospital, Beijing, China
guangleizhang@buaa.edu.cn

**Abstract.** The rapid on-site evaluation (ROSE) technique can significantly accelerate the diagnostic workflow of pancreatic cancer by immediately analyzing the fast-stained cytopathological images with on-site pathologists. Computer-aided diagnosis (CAD) using the deep learning method has the potential to solve the problem of insufficient pathology staffing. However, the cancerous patterns of ROSE images vary greatly between different samples, making the CAD task extremely challenging. Besides, due to different staining qualities and various types of acquisition devices, the ROSE images also have complicated perturbations in terms of color distribution, brightness, and contrast. To address these challenges, we proposed a novel multiple instance learning (MIL) approach using shuffle patches containing the instances, which adopts the patch-based learning strategy of Vision Transformers. With the re-grouped bags of shuffle instances and their bag-level soft labels, the approach utilizes a MIL head to make the model focus on the features from the pancreatic cancer cells, rather than that from various perturbations in ROSE images. Simultaneously, combined with a classification head, the model can effectively identify the general distributive patterns across different instances. The results demonstrate the significant improvements in the classification accuracy with more accurate attention regions, indicating that the diverse patterns of ROSE images are effectively extracted, and the complicated perturbations of ROSE images are significantly eliminated. It also suggests that the MIL with shuffle instances has great potential in the analysis of cytopathological images.

## 1 Introduction

Pancreatic cancer is one of the malignant tumors with the worst prognosis, whose 5-year survival rate is only 10% [1, 2]. Early diagnosis and treatment of pancreatic cancer can significantly improve the survival rate of patients. Endoscopic ultrasonography-guided fine-needle aspiration (EUS-FNA) is an important method for the pathological diagnosis of pancreatic cancer [3, 4]. Combined with EUS-FNA surgery, rapid on-site evaluation (ROSE) can quickly obtain cytopathological images to improve diagnostic efficiency and reduce the number of punctures during surgery, reducing pain and the risk of postoperative infection [5]. However, the broader application of the ROSE technique is limited by the shortage of pathologists, which calls for on-site AI to aid the workforce [6]. Computer-aided diagnosis (CAD) system shows high potential in solving this limitation [7-9].

At present, the research on CAD application for ROSE image classification is still in its infancy. In 2017, a multi-layer perceptron (MLP) model with an accuracy of 83.9% was established, based on the features extracted from the region of interest (ROI), such as cell contour and image intensity [10]. In 2018, Hashimoto et al. [11] firstly applied deep learning to analyze ROSE images without manual feature extraction. This work trained and tested a model with 450 pathological images, and the achieved sensitivity, specificity, and accuracy were all 80%. In 2020, the same group applied another deep learning strategy to the ROSE image classification and achieved 13% higher accuracy with 1440 images [12]. The studies proved that deep learning has important clinical value in cytopathological image analysis. However, the current works on ROSE image classification have paid little attention to the perturbation characteristics of images, and the interpretability can be further improved.

There are many challenges in the classification of ROSE images. Due to the influence of staining quality, the images often contain impurities, such as broken cells, pollution, fibers, light spots, and so on, which affects the AI-onsite classification. The difference in acquisition equipment also affects the image, creating differences among brightness, contrast, saturation, and resolution. The thickness of the slices brings difficulties in focusing, which means parts of the image are often vague, and the model should avoid the regional misleading [13]. Despite individual differences, the cytopathological characteristics of ROSE are also complex, making it difficult for the model to obtain a clear decision boundary towards cancerous patterns in various images.

Multiple instance learning (MIL) enables the model to focus on the differences among instances by bagging multiple instances and providing bag-level labels, showing potential in the classification task with complex background [14-16]. Based on MIL, we propose a novel shuffle strategy with data augmentation to keep the instances with different staining and acquisition qualities in the same bag. With this strategy, the model is made to focus on the differences among cells rather than various perturbations of ROSE images under the supervision of bag-level soft labels. In addition, with clinical knowledge, the differences of normal cells and cancer cells arise from local features (cell size, nucleocytoplasmic ratio, cell shape, nuclear membrane shape, chromatin distribution, etc.) and global features (overall orientation of cells, orientation of cell clusters and spacing between cells, etc.). Therefore, a combination of MIL head and classification (CLS) head can balance the feature extraction on the local and global features respectively. Combining the above strategies, we propose a multiple instance learning with shuffle instances (MIL-SI) approach to classify the pancreatic ROSE images.

## 2    Method

Towards the characteristics of the ROSE images, the MIL-SI approach is composed of two soft-label distributers, a Vision-Transformer-based backbone, and two task-based heads (a MIL head for soft label regression and a CLS head for bag label prediction). Two steps of the MIL and CLS are specially designed. In the MIL step, bags of shuffled instances are composed into images, and after the feature extraction of the backbone, the MIL head is used to regress the soft label of the bag. In the CLS step, with un-shuffled image patches sent as input bags, the MIL head and an additional CLS head are used to predict the categories. The approach is described in **Fig. 1.**

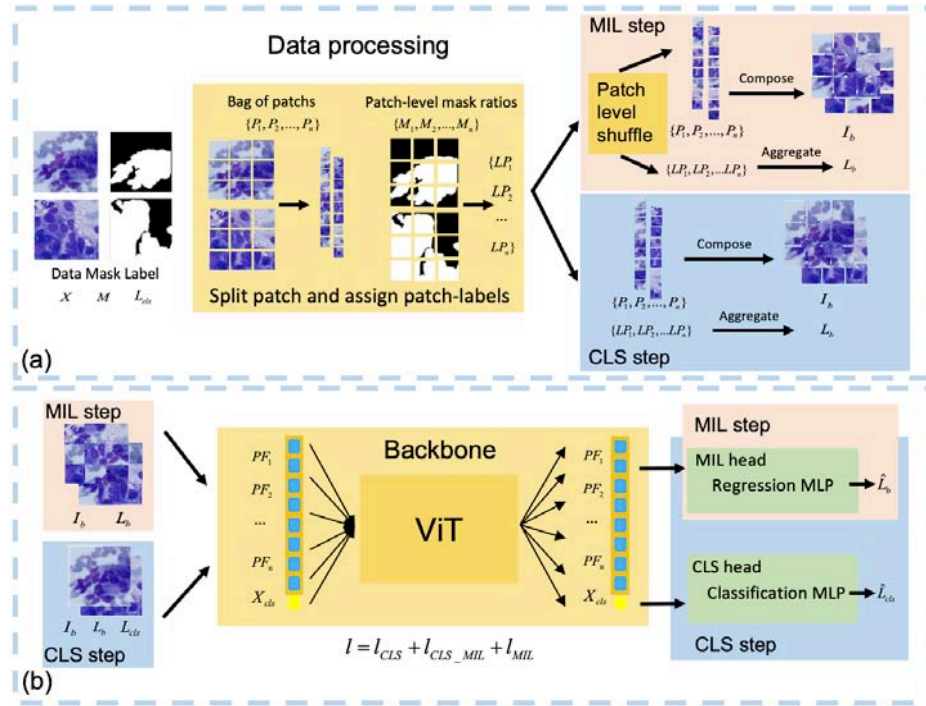

**Fig. 1.** Overview of MIL-SI composed of two steps: MIL step and CLS step. In the data processing as illustrated in (a), the images are transformed into patches, and the patch label is calculated based on the corresponding masks. In the MIL step, the bags of patches within a batch are shuffled while the bags of image patches remain unchanged in the CLS step. Lastly, the bags are composed of images with their soft label aggregated from the patch-level label. In the proposed two-step training process in (b), the backbone extracts the features of shuffled and un-shuffled images. Then, the patch tokens are used to regress the bag-level soft-label in the MIL head. In the CLS step, an additional CLS head is used to predict the categories of the input images based on the class token.

### 2.1 Multiple Instance Learning

Defined by Dietterich et al. [17], MIL models a bag of instances together and the bag label is related to the instance label. The formulation of bag label is

$$L = \begin{cases} 0, & if \ \sum L_i = 0 \\ 1, & otherwise \end{cases} \tag{1}$$

where $L$ is the bag label and $L_i$ is the label of instance $x_i$, $L, L_i \in [0,1]$, $i = 1, 2, ..., n$.

In many related works [18-20], MIL model can predict the bag label while the instance label is not accessible. The modeling process can be separated into two stages: instance level feature extraction by backbone and bag level modeling from features to label by MIL head, given by

$$\hat{L}(X) = MIL(backbone(X)) \tag{2}$$

where $X = \{x_1, x_2, ..., x_n\}$ is a bag of instances, $\hat{L}(X)$ is the prediction label of $X$.

In addition to the prediction of the discrete label, the bag-level tumor purity is regressed by MLP while the instance label is not accessible, showing that the regression MLP is applicable in the lightweight supervision on the whole bag [19].

### 2.2 Preprocessing and the Soft-label Distributers

**Preprocessing.** After augmentation of the ROSE images and their masks, each ROSE image $X$ and its mask $M$ are sequentially divided into patches $\{P_1, P_2, ..., P_n\}$ and mask patches $\{M_1, M_2, ..., M_n\}$. The patch size is critical as the scale may influence the cells in each patch. In the shuffling process, the cells will be cut into pieces if the patch size is less than 16 pixels, and therefore the spatial features may be lost. On the contrary, if the patch size is too large, the cell groups may be reserved, causing the shuffling operation to be less effective in reducing the perturbations of cells.

**Soft-label Distributers.** After the patch split operation, each patch $P_i$ of the image $X$ is assigned with a patch label to represent its masked area of categories. The patch label is defined as

$$LP_i = \{MR_i, MR_{1i}, MR_{2i}, ..., MR_{Ki}\} \tag{3}$$

where $LP_i$ is the label of patch $P_i$, $MR_i$ is the ratio of masked pixels in $P_i$, and $MR_{ji}$ is the ratio of masked pixels of category $j$, $j = 1, 2, ..., K$. $K$ is the number of categories. For each patch, $MR_{ji}$ is defined as

$$MR_{ji} = \begin{cases} MR_i, & if\ j = k \\ 0, & otherwise \end{cases} \tag{4}$$

where k is the category of $P_i$.

Then, we designed two soft label distributers: MIL distributer and CLS distributer. In the MIL soft label distributer, the patches of bags in a batch are randomly shuffled to obtain mixed bags, while the bags remain un-shuffled in the CLS soft label distributer. Lastly, both distributers compose the bag of instances into an image and calculate its bag-level soft label, and the Transformer backbone models features from the composed image $I_b$. To create pre-knowledge supervision for the MIL task, the bag-level soft label $L_b$ is given as

$$L_b = \{\sum MR_i, \sum MR_{1i}, \sum MR_{2i}, ..., \sum MR_{Ki}\}. \tag{5}$$

### 2.3 Vision Transformer Backbone

**Backbone ViT.** The Vision Transformers [21, 22] have achieved inspiring results on various computer vision tasks, whose patch-based learning strategy is suitable for the MIL-SI approach, so we adopt ViT as the backbone. The feature sequence after feature extraction of the backbone is

$$\{X_{cls}, PF_1, PF_2, ..., PF_n\} = backbone(Embedding(I_b)) \tag{6}$$

where $X_{cls} \in R^D, PF_i \in R^D$, $D$ is the embedding dimension, $X_{cls}$ is the CLS token, and $PF_i$ is the output token of backbone.

After the feature extraction, the CLS token $X_{cls}$ is used for the CLS head to predict the categories, and the rest of tokens $\{PF_1, PF_2, ..., PF_n\}$ representing the patch features are reserved for the MIL head.

Therefore, the prediction of bag-level soft label $\hat{L}_b$ and classification label $\hat{L}_{cls}$ are defined as

$$\hat{L}_b = MLP(\{PF_1, PF_2, ..., PF_n\}), \tag{7}$$

$$\hat{L}_{cls} = CLS(X_{cls}). \tag{8}$$

### 2.4 Two Task-based Heads and Dataflow

Adapt to the two labels, the MIL-SI model is designed with two heads: a CLS head and a MIL head, which predict the classification label and regress the bag-level soft label simultaneously. To our knowledge, this is the first work using the patch-level shuffle strategy for MIL to improve the model performance.

The dataflow of the MIL-SI approach is illustrated in **Fig. 1.** For each ROSE image considered as a bag within a batch, the image and its mask are divided into patches, and the soft label of each patch is assigned based on its mask. Secondly, two dataflows are specially designed. In the MIL step, the bags within a batch are shuffled on patch level randomly, so the cells from different patches are grouped in a new bag. On the contrary, in the CLS step, the bags are not shuffled to remain their global features. After the bag distribution, each bag of patches is composed back to an image with the correspond bag-level soft label calculated in the meantime, as input for the model.

Then, the heads and model are trained with two steps, both steps use the backbone to extract features. In the MIL step, the patch features extracted from the shuffled bag are modeled in the MIL head to regress the bag-level soft labels. In the CLS step, the MIL head is also used based on un-shuffled bags, and the CLS head is used to predict the classification label from the CLS token in the Transformer. Lastly, the loss is calculated by

$$l = l_{MIL} + l_{MIL\_CLS} + l_{CLS} \tag{9}$$

where $l_{MIL}$ is loss from MIL head in the MIL step, $l_{MIL\_CLS}$ is loss from MIL head in the CLS step and $l_{CLS}$ is loss from CLS head in the CLS step.

## 3 Experiment

### 3.1 Dataset and Experimental Details

**Dataset.** The EUS-FNA and ROSE diagnoses were performed in Peking Union Medical College Hospital, and the ROSE images were collected under the supervision of senior pathologists. The enrolled images were acquired by two microscope digital cameras (Olympus BX53 and Nikon Eclipse Ci-S) through the acquisition heads of Basler ScA1 and Olympus DP73. The images were saved in 'jpg' format with a resolution of 1390*1038. A total amount of 1773 pancreatic cancer images and 3315 normal pancreatic cell images were collected, and the classification labels and segmentation results were confirmed by senior pathologists. The whole 5088 images were divided into a training set, validation set, and test set at the ratio of 7:1:2.

**Implementation Details.** The same data augmentation strategy was designed in each experiment to deal with the data scarcity and image perturbations. In the data augmentation process, the data and mask were randomly rotated, and then the center area of 700*700 pixels was reserved. Secondly, the area was resized to 384*384 pixels, and the random vertical and horizontal flip were applied. Lastly, for each ROSE image, the PyTorch ColorJitter (with brightness = 0.15, contrast = 0.3, saturation = 0.3, and hue = 0.06) was used to recreate the perturbations during the sampling process. In the validation and test processes, only the center-cut and the resizing operations were applied, obtaining the center area of 700*700 pixels and resizing it to 384*384 pixels.

The model was trained for 50 epochs, and the model with the highest validation accuracy was saved as the output model. The Adam optimizer [23] was applied with a learning rate of e-5 and a momentum of 0.05. The cosine learning rate decay strategy was adopted to reduce the learning rate twenty times sequentially in the training process. The same training parameters were used on all experiments.

The experiments were carried out and recorded online on the Google CoLab pro+ platform. As a lightweight approach, a single 16 GB Nvidia P100- PCIe GPU was used with Python version 3.7.12 and Pytorch version 1.9.0+cu111.

**Baselines.** To compare the performance of our proposed MIL-SI on the ROSE classification task, several widely used CNNs and Transformers were compared as counterpart models. And to prove the effectiveness of the shuffle instances strategy, the MIL model using only CLS step (with the MIL head working on the un-shuffled bag) and the backbone model (ViT) were compared.

The models were built based on the timm library, and the transfer learning strategy was applied to all models with their official weights.

**Metrics.** Accuracy, precision, recall, specificity, and F1-score were used to evaluate the performance of models.

### 3.2 Experimental Results

To verify the effectiveness of our approach, we compared MIL-SI with other models under the same experimental conditions. As shown in **Table 1**, MIL-SI achieves the best results among these models in accuracy, precision, recall, etc. Compared with other models, the accuracy of MIL-SI is improved by 3.2%-4.8%, and in other indicators, especially precision and recall are improved by approximately 6%. In addition, by visualizing the gradient-weighted classification activation mapping (Grad-CAM), our model obtains the most accurate attention regions compared with other models (**Fig. 2**). Both in positive and negative images, our model can accurately focus on the areas of the cell clusters. Therefore, MIL-SI pays more attention to the pancreatic cells by shuffle instances strategy, rather than over-fitting the background perturbations.

**Table 1.** Performance of Models on ROSE Image Classification

| Model | Accuracy (%) | Precision (%) | Recall (%) | Specificity (%) | F1-score (%) |
|---|---|---|---|---|---|
| Efficient Net [24] | 89.57 | 85.03 | 85.03 | 91.99 | 85.03 |
| Inception v3 [25] | 90.75 | 86.72 | 86.72 | 92.90 | 86.72 |
| ResNet50 [26] | 90.75 | 87.36 | 85.88 | 93.35 | 86.61 |
| Swin Transformer [22] | 89.17 | 86.75 | 81.36 | 93.35 | 83.97 |
| ViT [21] | 90.65 | 88.20 | 84.46 | 93.96 | 86.29 |
| MIL-SI | **94.00** | **91.98** | **90.68** | **95.77** | **91.32** |

**Table 2.** Comparison of Shuffle Strategy and Patch Size on MIL-SI

| Model | Patch size | Accuracy (%) | Precision (%) | Recall (%) | Specificity (%) | F1-score (%) |
|---|---|---|---|---|---|---|
| ViT | - | 90.65 | 88.20 | 84.46 | 93.96 | 86.29 |
| MIL-CLS | 32 | 92.13 | 90.06 | 87.01 | 94.86 | 88.51 |
| MIL-SI | 32 | **94.00** | **91.98** | **90.68** | **95.77** | **91.32** |
| MIL-SI | 16 | 93.60 | 92.88 | 88.42 | 96.37 | 90.59 |
| MIL-SI | 32 | **94.00** | 91.98 | **90.68** | 95.77 | **91.32** |
| MIL-SI | 64 | 93.11 | 91.04 | 88.98 | 95.32 | 90.00 |
| MIL-SI | 128 | 92.62 | **92.92** | 85.31 | **96.53** | 88.95 |

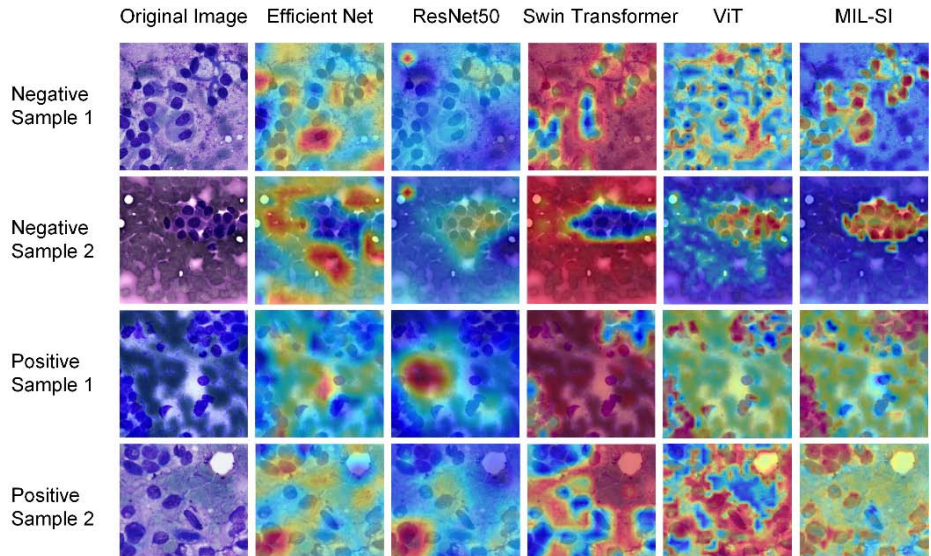

**Fig. 2.** The Grad-CAMs of Models on ROSE Image Classification

To verify the effectiveness of the shuffle instances strategy, we conducted the MIL-CLS model without the shuffle operation and compared MIL-SI with it in addition to the naive ViT. As shown in **Table 2**, the introduction of the soft label and MIL head can improve the classification accuracy significantly (90.65% to 92.13%) even without the shuffle operation, the model learns global features with the un-shuffled bag in the CLS step simultaneously. The MIL-SI achieved the highest results with the shuffle strategy, indicating that the model pays more attention to the differences among pancreatic cells by reducing perturbations. Besides, we tested the effect of patch size and found that the best size is 32, which means each patch contains 1-2 cells, and therefore the local features of cells are reserved.

## 4      Conclusion

This paper firstly proposed a novel shuffle instances-based MIL approach in ROSE image classification. With the patch shuffling strategy, the MIL-SI has the cell instances within a batch shuffled and re-grouped into bags. The proposed MIL-SI can effectively reduce the complicated perturbations among ROSE images and enhance the model to concentrate on cell features. The MIL-SI simultaneously extracts local features and models the general distributive patterns across different cells by combining the MIL head with the classification head. Towards the common challenges in cytopathological image analysis, the MIL-SI utilizes a straightforward strategy to balance the various perturbations, complicated spatial features, and general distributive patterns. The inspiring experimental results illustrate the improvements in the modeling and robust ability of the model, showing great potential in cytopathological image analysis.

**References**

1. Siegel R.L., Miller K.D., Fuchs H.E., and Jemal A., Cancer Statistics, 2021. Ca-a Cancer Journal for Clinicians 71 (2021) 7-33. http://doi.org/10.3322/caac.21654
2. Zeng H.M., Chen W.Q., Zheng R.S., Zhang S.W., Ji J.S., Zou X.N., et al., Changing cancer survival in China during 2003-15: a pooled analysis of 17 population-based cancer registries. Lancet Global Health 6 (2018) E555-E567. http://doi.org/10.1016/s2214-109x(18)30127-x
3. Facciorusso A., Martina M., Buccino R.V., Nacchiero M.C., and Muscatiello N., Diagnostic accuracy of fine-needle aspiration of solid pancreatic lesions guided by endoscopic ultrasound elastography. Annals of gastroenterology 31 (2018) 513-518. http://doi.org/10.20524/aog.2018.0271
4. Matsubayashi H., Sasaki K., Ono S., Abe M., Ishiwatari H., Fukutomi A., et al., Pathological and Molecular Aspects to Improve Endoscopic Ultrasonography-Guided Fine-Needle Aspiration From Solid Pancreatic Lesions. Pancreas 47 (2018) 163-172. http://doi.org/10.1097/mpa.0000000000000986
5. Yang F., Liu E., and Sun S., Rapid on-site evaluation (ROSE) with EUS-FNA: The ROSE looks beautiful. Endoscopic Ultrasound 8 (2019) 283-287. http://doi.org/10.4103/eus.eus_65_19
6. van Riet P.A., Cahen D.L., Poley J.-W., and Bruno M.J., Mapping international practice patterns in EUS-guided tissue sampling: outcome of a global survey. Endoscopy international open 4 (2016) E360-70. http://doi.org/10.1055/s-0042-101023
7. Calderaro J., and Kather J.N., Artificial intelligence-based pathology for gastrointestinal and hepatobiliary cancers. Gut 70 (2021) 1183-1193. http://doi.org/10.1136/gutjnl-2020-322880
8. Landau M.S., and Pantanowitz L., Artificial intelligence in cytopathology: a review of the literature and overview of commercial landscape. Journal of the American Society of Cytopathology 8 (2019) 230-241. http://doi.org/10.1016/j.jasc.2019.03.003
9. Marya N.B., Powers P.D., Chari S.T., Gleeson F.C., Leggett C.L., Abu Dayyeh B.K., et al., Utilisation of artificial intelligence for the development of an EUS-convolutional neural network model trained to enhance the diagnosis of autoimmune pancreatitis. Gut 70 (2021) 1335-1344. http://doi.org/10.1136/gutjnl-2020-322821
10. Momeni-Boroujeni A., Yousefi E., and Somma J., Computer-assisted cytologic diagnosis in pancreatic FNA: An application of neural networks to image analysis. Cancer Cytopathology 125 (2017) 926-933. http://doi.org/10.1002/cncy.21915
11. Hashimoto Y., Ohno I., Imaoka H., Takahashi H., Mitsunaga S., Sasaki M., et al., Reliminary result of computer aided diagnosis (cad) performance using deep learning in eus-fna cytology of pancreatic cancer. Gastrointestinal Endoscopy 87 (2018) AB434-AB434

12. Hashimoto Y., Prospective comparison study of EUS-FNA onsite cytology diagnosis by pathologist versus our designed deep learning algorhythm in suspected pancreatic cancer. Gastroenterology 158 (2020) S17-S17
13. Pantanowitz L., and Bui M.M., Image analysis in cytopathology. Modern Techniques in Cytopathology 25 (2020) 91-98. http://doi.org/10.1159/000455776
14. Butke J., Frick T., Roghmann F., El-Mashtoly S.F., Gerwert K., and Mosig A., End-to-end Multiple Instance Learning for Whole-Slide Cytopathology of Urothelial Carcinoma, MICCAI Workshop on Computational Pathology, PMLR, 2021, pp. 57-68
15. Li H., Yang F., Zhao Y., Xing X.H., Zhang J., Gao M.X., et al., DT-MIL: Deformable Transformer for Multi-instance Learning on Histopathological Image, International Conference on Medical Image Computing and Computer Assisted Intervention (MICCAI), Electr Network, 2021, pp. 206-216. http://doi.org/10.1007/978-3-030-87237-3_20
16. Yu S., Ma K., Bi Q., Bian C., Ning M.N., He N.J., et al., MIL-VT: Multiple Instance Learning Enhanced Vision Transformer for Fundus Image Classification, International Conference on Medical Image Computing and Computer Assisted Intervention (MICCAI), Electr Network, 2021, pp. 45-54. http://doi.org/10.1007/978-3-030-87237-3_5
17. Dietterich T.G., Lathrop R.H., and Lozano-Pérez T., Solving the multiple instance problem with axis-parallel rectangles. Artificial intelligence 89 (1997) 31-71
18. Chikontwe P., Kim M., Nam S.J., Go H., and Park S.H., Multiple instance learning with center embeddings for histopathology classification, International Conference on Medical Image Computing and Computer-Assisted Intervention, Springer, 2020, pp. 519-528
19. Oner M.U., Chen J., Revkov E., James A., Heng S.Y., Kaya A.N., et al., Obtaining spatially resolved tumor purity maps using deep multiple instance learning in a pan-cancer study. Patterns (2021) 100399
20. Shao Z., Bian H., Chen Y., Wang Y., Zhang J., and Ji X., Transmil: Transformer based correlated multiple instance learning for whole slide image classification. Advances in Neural Information Processing Systems 34 (2021)
21. Dosovitskiy A., Beyer L., Kolesnikov A., Weissenborn D., Zhai X., Unterthiner T., et al., An image is worth 16x16 words: Transformers for image recognition at scale. arXiv preprint arXiv:2010.11929 (2020)
22. Liu Z., Lin Y., Cao Y., Hu H., Wei Y., Zhang Z., et al., Swin transformer: Hierarchical vision transformer using shifted windows, Proceedings of the IEEE/CVF International Conference on Computer Vision, 2021, pp. 10012-10022
23. Kingma D.P., and Ba J., Adam: A method for stochastic optimization. arXiv preprint arXiv:1412.6980 (2014)
24. Tan M., and Le Q., Efficientnet: Rethinking model scaling for convolutional neural networks, International conference on machine learning, PMLR, 2019, pp. 6105-6114
25. Szegedy C., Vanhoucke V., Ioffe S., Shlens J., and Wojna Z., Rethinking the inception architecture for computer vision, Proceedings of the IEEE conference on computer vision and pattern recognition, 2016, pp. 2818-2826
26. He K., Zhang X., Ren S., and Sun J., Deep residual learning for image recognition, Proceedings of the IEEE conference on computer vision and pattern recognition, 2016, pp. 770-778